\documentclass[preprint,12pt]{elsarticle}
 \biboptions{authoryear,sort}

 \usepackage{graphicx}
\usepackage{epstopdf}
\usepackage{color}

\usepackage{amssymb}

\newcommand{\Jmax}{J_\mathrm{max}}
\newcommand{\Jmin}{J_\mathrm{min}}

\newcommand{\lsim}{\lesssim\!}

\def\apj{ApJ}
\def\mnras{MNRAS}

\def\araa{ARA\&A}                
\def\apjl{ApJ}                   

\def\prd{Phys. Rev. D}

\def\ssr{Space Sci. Rev.}
\def\aapr{Astron. Astroph. Reviews}

\newcommand{\xx}[1]{\!\times\!10^{#1}}

\newcommand{\ACSF}{A_\mathrm{CSF}}
\newcommand{\Atot}{A_\mathrm{tot}}
\newcommand{\FEx}{F_\mathrm{ex}}
\newcommand{\FGal}{F_\mathrm{gal}}
\newcommand{\VxMean}{\overline{v_{x}}}
\newcommand{\VyMean}{\overline{v_{y}}}
\newcommand{\VzMean}{\overline{v_{z}}}
\newcommand{\uZp}{u_{z'}}
\newcommand{\uMean}{\overline{u_{z'}}}
\newcommand{\xMean}{\overline{u_{x'}}}
\newcommand{\yMean}{\overline{u_{y'}}}

\newcommand{\fEx}{f_\mathrm{ex}}

\newcommand{\Msun}{M_\odot}

\newcommand{\EnCR}{E_\mathrm{CR}}

\newcommand{\MW}{Milky Way}

\newcommand{\delMax}{\Delta\theta_\mathrm{max}}

\newcommand{\muG}{$\mu$G}

\newcommand{\rel}{relativistic}

\newcommand{\mc}{Monte Carlo}

\newcommand{\Facc}{Fermi acceleration}
\newcommand{\FoFSA}{first-order Fermi shock acceleration}

\usepackage[ps2pdf,%
a4paper=true,%
breaklinks=true,%
colorlinks=true,%
pdfauthor={First Author et al.},%
pdftitle={Template for manuscripts in Advances in Space Research}%
]{hyperref}

\journal{Advances in Space Research}

\begin{document}

\begin{frontmatter}



\title{High-energy cosmic rays from compact galactic star clusters: particle fluxes and anisotropy}




\author{Bykov A.M.$^{1}$, Kalyashova M.E.$^{1}$, Ellison D.C.$^{2}$ and Osipov S.M.$^{1}$}
\address{$^{1}$Ioffe Institute, Saint-Petersburg, Polytechnicheskaya str., 26, 194021, Russia}
\ead{byk@astro.ioffe.ru, filter-happiness@yandex.ru,ellison@ncsu.edu,osm2004@mail.ru}
%
\address{$^{2}$North Carolina State University, Department of Physics, Raleigh, NC 27695-8202, USA}

\begin{abstract}
It has been shown that supernova blast waves interacting with winds from massive stars in compact star clusters may be capable of 
producing cosmic-ray (CR) protons to above $10^{17}$\,eV. 
We give a brief description of the colliding-shock-flows mechanism and look at generalizations of the diffusion of $\sim 100$\,PeV CRs in the turbulent galactic magnetic field present in the galactic disk.
We calculate the temporal evolution of the CR anisotropy from a possible distribution of young compact massive  star clusters
assuming the sources are intermittent on time scales of a few million years, i.e., comparable to their residence time in 
the \MW. 
Within the confines of our model, we determine the galactic/extra-galactic fraction  of high-energy CRs resulting in anisotropies consistent with observed values.
We find that galactic star clusters may contribute a substantial fraction of $\sim 100$\,PeV CRs without producing anisotropies above observed limits. \\
\end{abstract}


\end{frontmatter}
\section{Introduction}
The main source of galactic cosmic rays (CRs) with energies in the TeV energy regime is almost certainly supernova remnants (SNRs) which occur on time scales which are short compared to the CR residence time in the disk 
\citep[see e.g.][]{Berezinski90,BaringEtal99,2012SSRv..166...97A,SMP2007,2012APh....39...52D,helderea12,amato14,blandford14,2015MNRAS.447.2224B,2015ARA&A..53..199G}.
The source of CRs with energies in the PeV-EeV range is far less certain. Isolated SNRs with typically observed characteristics are unlikely to produce PeV-EeV CRs although \rel\ supernovae (SNe) may be able to produce CRs well above the ``knee" region at $10^{15-16}$\,eV 
\citep[e.g.,][]{2007PhRvD..76h3009W,Chakraborti2011,2018SSRv..214...41B}.
The energy budget requirement for the population of CRs above the knee is less stringent than for lower energy CRs and the sources of 
PeV-EeV CRs may be fundamentally different from isolated SNRs. 
Here we consider CRs produced in colliding shock flows (CSFs) in galactic compact star clusters as proposed by \citet{Bykov2014,Bykov2018ASR}.

\begin{figure}[h]
\centering
\includegraphics[width=10 cm]{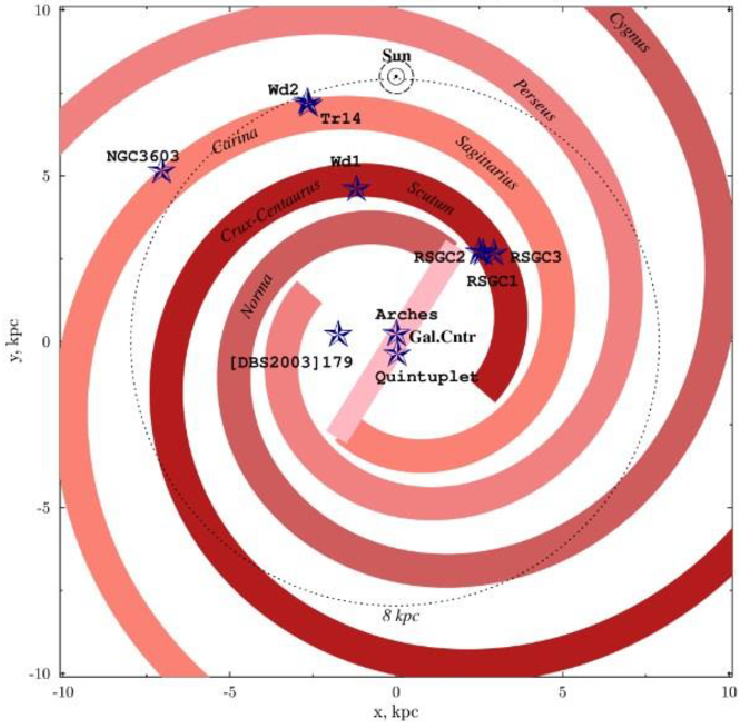}         
\caption{Positions  of some young massive star clusters observed in the Galaxy from \citet{Portegies_Zwart2010}.   
\label{fig:clusters}}
\end{figure}

\begin{figure}[h]
\centering
\includegraphics[width=10 cm]{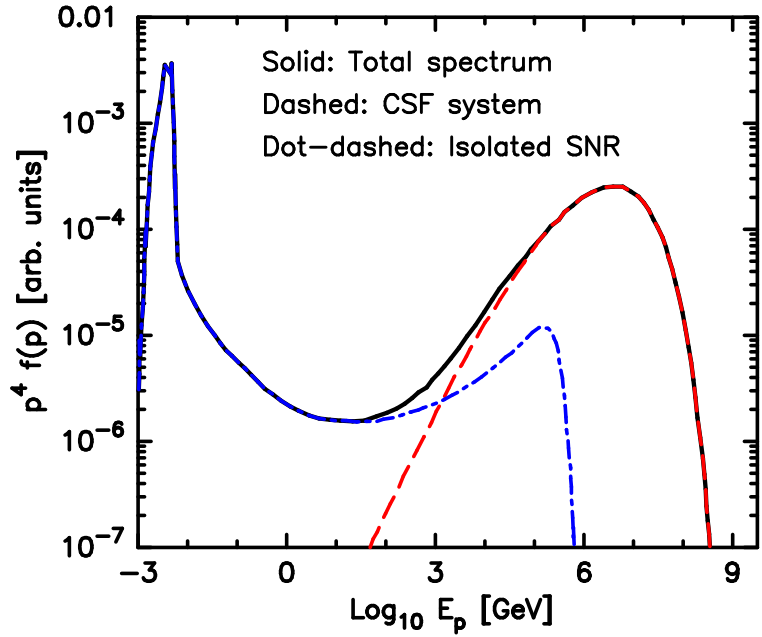}           
\caption{Cosmic ray protons obtained from a CSFs model as described in 
\citet{Bykov2018ASR}. The dashed (red) curve shows protons from the stage when the SNR shock is interacting with a fast stellar wind. The dot-dashed (blue) curve shows the spectrum from an isolated SNR, and the solid (black) curve is the total.  
\label{fig:spec}}
\end{figure}

While many SNe may occur in isolation, it's clear that SNe also occur in compact star clusters including clusters with masses $>10^5\Msun$. These massive clusters may contain thousands of young massive  stars within the few parsec scale size of the cluster.
The most massive stars will explode within a few million years from the birth of the cluster and it has been shown 
\citep[e.g.][]{Bykov2014,BEGO2015MNRAS,2018AdSpR..62.2750L,Grimaldo2019} that, if the blast waves from these core-collapse SNe interact with the strong stellar winds from nearby massive stars,
efficient \Facc\ can produce protons to hundreds of PeV  with an extremely hard spectrum near the cutoff at the maximum obtained energy 
\citep[see Fig.~\ref{fig:spec} from][]{Bykov2018ASR}.

Furthermore, if the \FoFSA\ mechanism accelerates heavy ions in CSF systems in the same way as believed to be the case in other shocks \citep[e.g.,][]{EDM97},  the high-energy peak shown in Fig.~\ref{fig:spec} will be shifted upward by a factor equal to the CR charge. Iron nuclei might be accelerated to EeV energies.

\cite{factories18} derived an approximate $1/r$ decrement of the CR density with distance from a star cluster from the analysis of very high energy gamma-ray observations of  massive star clusters like  Westerlund 1 and 2, the Cygnus cocoon, and others. 
This may indicate a continuous high energy CR injection into the interstellar medium over a few million years. 
They pointed out that such sources may operate up to the CR knee 
around 1 PeV.  This is consistent with the modeling of high energy CR spectra in compact clusters of young massive stars by 
\citet[][]{Bykov2014,BEGO2015MNRAS,Bykov2018ASR}  discussed above.  
A critical constraint on the possible contribution of galactic sources to PeV-EeV CRs is the low observed anisotropy. Here we calculate CR diffusion in the \MW\ with a simulated turbulent magnetic field, 
as described in \S~\ref{sec:Bfield}.
The CRs are injected from sources mimicking the known massive star clusters and are propagated using a \mc\ method described in \S~\ref{sec:MCprop}.

\section{Model}

\subsection{Magnetic field and diffusion coefficients} \label{sec:Bfield}
\begin{figure}[h!]
\centering
\includegraphics[width=10 cm]{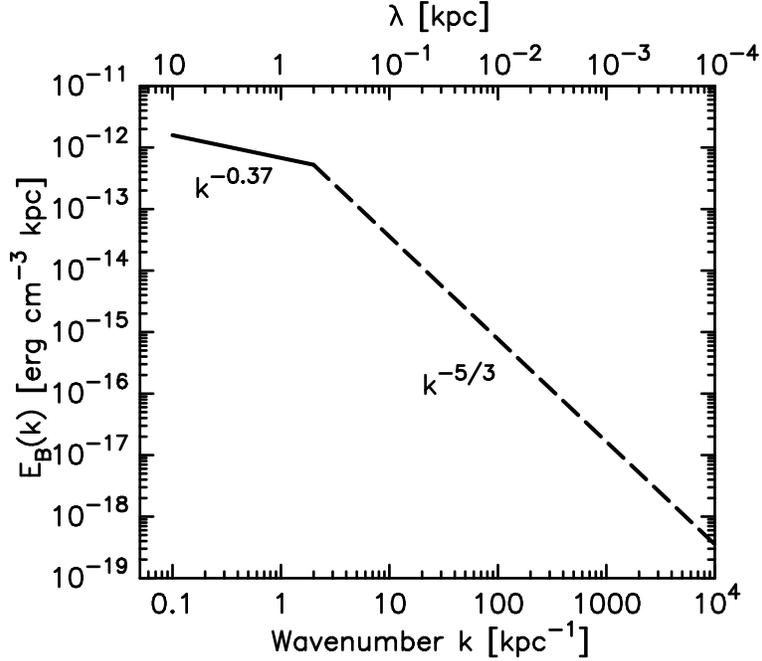}           
\caption{The energy spectrum of the turbulent magnetic field compiled by \citet{Han2017} from the analysis of rotation measures and dispersion measures of pulsars. The spectrum is used to simulate the diffusion coefficient of high energy CRs propagating 
in the Galaxy shown in Fig. \ref{fig:diff1}.    
\label{fig:HanSpec}}
\end{figure}

To calculate the galactic magnetic field, we use the most recent model of the turbulent magnetic field suggested by \citet{Han2017}. 
In this model, it is considered that on small
spatial scales both density and magnetic field 
fluctuations follow a Kolmogorov spectrum, as shown with the dashed line in Fig.~\ref{fig:HanSpec}.
By using rotation measures (RMs) and dispersion measures (DMs) of pulsars, \citeauthor{Han2017} derived the spatial energy spectrum of the Galactic interstellar magnetic field at scales from
$1/k = 0.5$ to $15$\,kpc, where $k$ is the wavenumber. It was found that the spectrum in this scale range is much flatter than the Kolmogorov
spectrum extended to small scales and the turnover occurs between $0.5$\,kpc and 
$80$\,pc. 
Taking this into account, we use the spectrum shown in 
Fig.~\ref{fig:HanSpec} for our modeling.

We model the turbulent magnetic field as a sum of plain waves with random directions, phases, and polarizations \citep[e.g.,][]{GJ1994, Casse2002}, i.e.,
\begin{equation}
\mathbf{B}_{turb}({\bf r}) = \sum^N_{n=1} A_n e^{i({\bf k_n \cdot r} + \psi_n)} \hat{\xi}_n
\ ,
\end{equation}
where $\mathbf{r}$ is the position vector, $\mathbf{k}_n \equiv k_n \hat{\mathbf{e}}^1_n$, $A_n$, $\psi_n$ and $\hat{\xi}_n$ are the wavevector, amplitude, phase and polarization vector of each mode, respectively. 
The polarization vector is given by:
\begin{equation}
    \hat{\xi}_n = \cos(\beta_n)\hat{\mathbf{e}}^2_n+i\sin(\beta_n)\hat{\mathbf{e}}^3_n
\ ,
\end{equation} 
where $\beta_n$ is the polarization angle. $(\hat{\mathbf{e}}^1_n, \hat{\mathbf{e}}^2_n, \hat{\mathbf{e}}^3_n)$  are orthogonal, so $\mathbf{k_n} \cdot \hat{\xi}_n = 0$ and 
$\nabla \cdot \mathbf{B}_\mathrm{turb}=0$.
The amplitude of each mode is determined with the energy spectrum shown in Fig.~\ref{fig:HanSpec}  and the root-mean-square field is $6$\,\muG.
We also discuss below a possible effect of the structure of the regular large-scale galactic magnetic field
on the anisotropy of $\sim 10^{17}$\,eV CRs accelerated by galactic supernovae in the compact star clusters.

\begin{figure}[h!]
\centering
\includegraphics[width=10 cm]{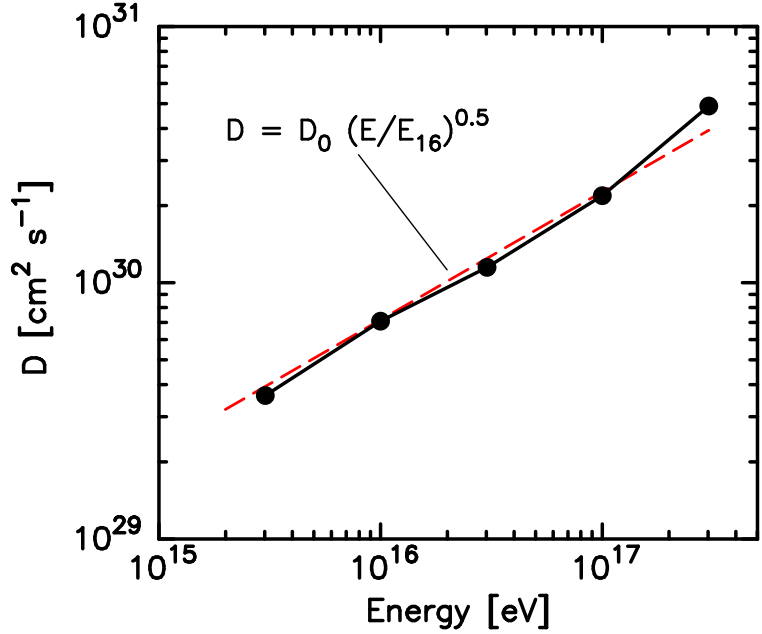}           
\caption{Diffusion coefficient in simulated turbulent magnetic fields 
having the spectrum shown in Fig.~\ref{fig:HanSpec}. The dashed red curve is a power law with $D_0=7\xx{29}$\,cm$^2$-s$^{-1}$ and $E_{16}$ is the CR energy in units of $10^{16}$\,eV. 
\label{fig:diff1}}
\end{figure}

\subsection{Cosmic-ray anisotropy}
To determine the dipole anisotropy at a given angle $\Theta$, we assume the CR
intensity $J(\mu)$ is
\begin{equation}
J(\mu) = J_0 + J_1\mu
\end{equation}
where $\mu= \cos \Theta$. The maximum intensity occurs when $\mu=1$ or $\Theta=0$. 
The dipole anisotropy, $|A|$, is defined as the maximum and minimum intensity
contrast, i.e.,
\begin{equation} \label{eq:Aniso}
|A| = \frac{\Jmax - \Jmin}{\Jmax + \Jmin} = 
\left | \frac{J_1}{J_0}\right |
\ .
\end{equation}
We note that eqn.~(\ref{eq:Aniso}) only applies when $J_1 \ll J_0$, i.e., when $|A| \ll 1$. The Auger CR observatory, and other experiments  
\citep[e.g.,][]{GoraAuger2018,ApelEtal2019}, show that the CR anisotropy at energies of $10^{17}$\,eV and above does not 
exceed $3-5$\% so eqn.~(\ref{eq:Aniso}) applies.

In a coordinate system where $z'$ is the direction of the 
maximum of intensity $I(\mu)$, and therefore the direction of the mean velocity,
the mean velocity $\uMean$ is
\begin{equation}
\uMean = \frac{\int \uZp J(\mu) d^3p}{\int J(\mu) d^3p} =
\frac{\int^1_{-1} cJ_1 \mu^2 d(\mu)}
{\int^1_{-1} J_0d(\mu)} =
\frac{cJ_1}{3J_0}
\ ,
\end{equation}
and
\begin{equation}
\xMean = \yMean = 0
\ .
\end{equation}
In the above, we assume the CR velocity $u\simeq c$ so $|A| \simeq |3\uMean/c|$. In the coordinates of our galaxy model, the $x$-$y$ plane is the galactic plane, the $z$-axis is directed upward, and 
$|\uMean| =\sqrt{\VxMean^2 + \VyMean^2 + \VzMean^2}$.

\subsection{\mc\ cosmic-ray propagation} \label{sec:MCprop}
Given the simulated turbulence discussed in \S~\ref{sec:Bfield}, we model the particle propagation by solving the equation of motion for a set of particles as they move through the turbulent magnetic field. From the motion, we find the diffusion coefficient, $D$, using
\begin{equation}
D = \frac{\left < \Delta r\right >^2}{6\Delta t}
\ ,
\end{equation}
where $\left < \Delta r\right >$ is the mean displacement of particles during the time $\Delta t$. Our results for CR energies $3\xx{15} - 3\xx{17}$\,eV  are shown
in Fig.~\ref{fig:diff1}. 

Once the diffusion coefficient is determined, the scattering mean-free-path is given by
\begin{equation} \label{eq:mfp}
\lambda = 3 D/c
\ ,
\end{equation}
and we run the \mc\ simulation with the following parameters. We follow two CR energies, $10^{17}$ and $3\xx{17}$\,eV, in a galaxy of 30\,kpc diameter and 10\,kpc thickness.
We assume supernova events occur every 2000\,yr and each source ejects $10^7$ \mc\ particles in random directions. 
The  particles undergo
large-angle scattering with the mean-free-path determined by 
eqn.~(\ref{eq:mfp}). 
The maximum angle of scattering is set as:
\begin{equation}
\delMax = \pi 
\ ,
\end{equation}
and the time step is 
\begin{equation}
\Delta t = 2 \lambda c
\ .
\end{equation}

At the position of the Earth, we find the 
anisotropy and particle flux in a small sphere of radius 0.5\,kpc
centered at the Solar System.
We follow each particle for a simulation time of $4 \xx {6}$\,yr or until it leaves the Galaxy. 
At every point in time, each particle will have coordinates $x$, $y$, and $z$ and velocity components $v_x$, $v_y$, and $v_z$.
With this information, we determine the CR concentration and 
anisotropy near Earth at each time step. 
The components of anisotropy are given by:
\begin{displaymath}
A_x = 3 \frac{\VxMean}{c} = \nonumber \\
3 \frac{\sum^N_{i=1} v_x^i/c}{N} 
\ ,
\end{displaymath}
\begin{equation}
A_y = 3 \frac{\VyMean}{c} =
3 \frac{\sum^N_{i=1} v_y^i/c}{N} 
\ ,
\end{equation}
\begin{displaymath}
A_z = 3 \frac{\VzMean}{c} =
3 \frac{\sum^N_{i=1} v_z^i/c}{N} 
\ ,
\end{displaymath}
where $N$ is the number of CRs in our region.
The total anisotropy is then $|A| = \sqrt{A_x^2 + A_y^2 + A_z^2}$.

\subsection{CR anisotropy from CSF sources}
In Fig.~\ref{fig:clusters} we show the distribution of 
known star clusters from \citet{Portegies_Zwart2010}.
We propagate CRs from these clusters to the Earth taking into account the large-scale structure of the Galactic magnetic field.
There are several models of the large-scale regular field  
\citep[see e.g.][]{Pshirkov,Jansson1,Han2017}. Here we use the 21-parameter model by  \citeauthor{Jansson1}. 
In our Monte Carlo simulations, CRs spiral in the large-scale regular field while scattering in the small-scale turbulence.

\begin{figure}[h]
\centering
\includegraphics[width=13.5 cm]{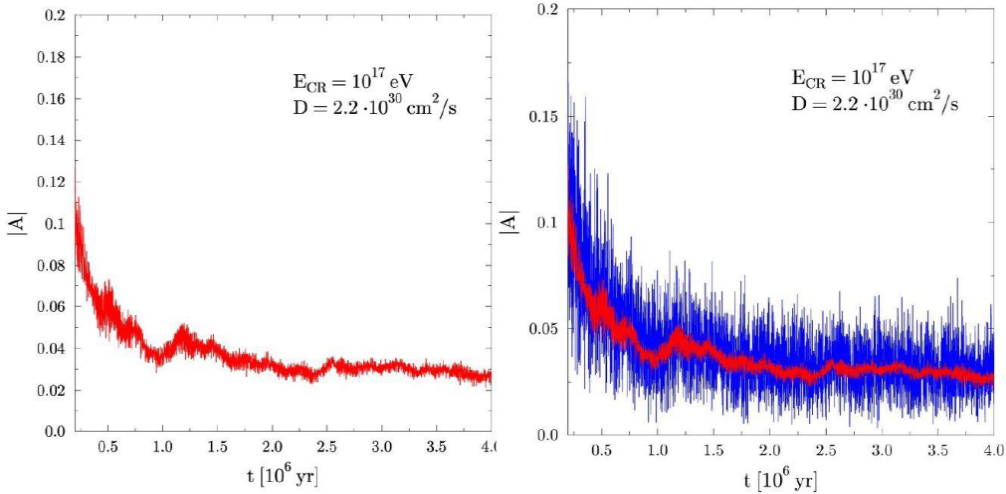}         
\caption{Anisotropy vs. time for 100 PeV protons.
The red curve in the left panel is for our CR scattering model without the regular Galactic magnetic field. In the right panel, the blue curve shows the effect of including the regular galactic field of \citet{Jansson1}. The poor statistics are due to  computational restrictions and the red curve is the same in both panels.
\label{fig:anis_03mar}}
\end{figure}

In  Fig.~\ref{fig:anis_03mar} 
we show the simulated anisotropy  for $10^{17}$\,eV CRs. 
Our simulations with the parameterized regular magnetic field are very computationally demanding, therefore, these simulations had only 10$^5$  particles per SNe compared to $10^7$ particles for the model without the regular magnetic field shown in the left panel in Fig.~\ref{fig:anis_03mar}.
This results in the higher noise level of the anisotropy (blue line in the right panel in Fig.~\ref{fig:anis_03mar}). 
The time history shows that $|A|$ reaches nearly constant minimum
values after $\sim 10^6$\,yr. 
As  seen in the right panel, the mean anisotropy is similar with and without propagation in the regular field of 
\citet{Jansson1}. 

%

\begin{figure}
\centering
    \includegraphics[scale=0.6]{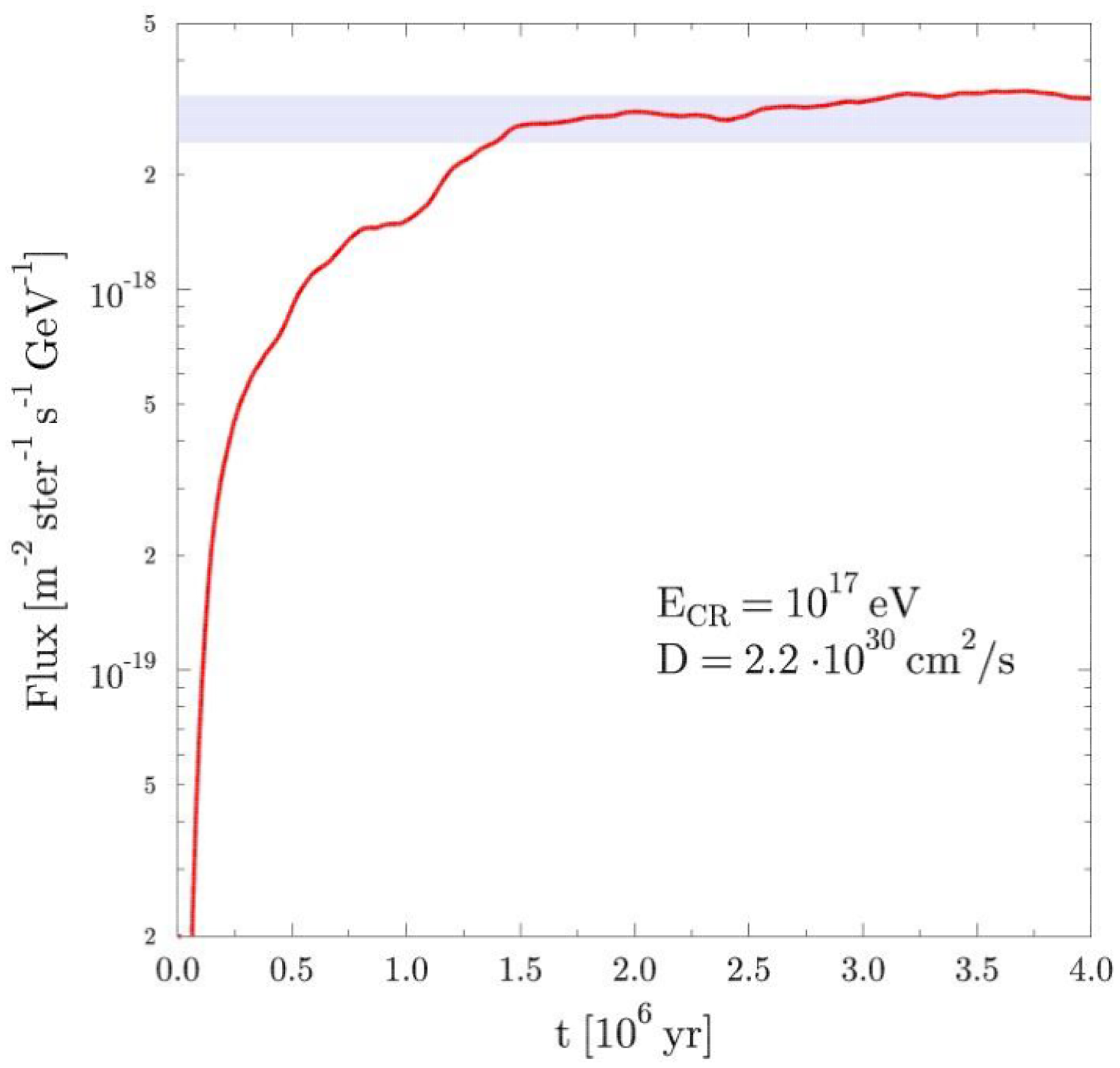}   
     \caption{Flux at Earth assuming the efficiency of converting of the SN energy to the kinetic energy of 100 PeV protons is $\chi=0.015 \%$. The pale blue band corresponds to the observed $10^{17}$\,eV CR flux near Earth. }
     \label{fig:flux}
\end{figure}

Fig.~\ref{fig:flux} shows the simulated flux at Earth as a function of run time, assuming that the efficiency of 
converting the SN explosion energy ($10^{51}$\,erg) to the energy of 100 PeV protons is about $0.015\%$. The presence of the regular magnetic field doesn't make any significant change in the flux which becomes 
nearly constant after $\sim 1.5\xx{6}$\,yr.

The observed equatorial dipole anisotropy at a few times $10^{17}$\,eV is $\lsim 0.01$ 
\citep[e.g.,][]{Mollerach2018}. For the parameters used here, 
we obtain $|A| \simeq 0.03$ 
for $\EnCR=10^{17}$\,eV.
An isotropic extra-galactic component of CRs will reduce the 
anisotropy and we can estimate the ratio of the galactic to extra-galactic flux, $\FGal/\FEx$, in the following way.
If the isotropic extra-galactic flux is $\fEx$ times as large as the average flux from the CSF sources, the total anisotropy is
\begin{equation}
\Atot = \frac{\ACSF}{1 + \fEx} 
\ ,
\end{equation}
and
\begin{equation}
\frac{\FGal}{\FEx} = \frac{1}{\fEx} =
\frac{\Atot}{\ACSF - \Atot}
\ .
\end{equation}
Taking $\Atot=0.01$, our results suggest that $\FGal/\FEx \sim 1/2$ for $\EnCR=10^{17}$\,eV.

\section{Conclusions}
As shown in \citet{Bykov2018ASR}, and references therein, the colliding shock flows (CSFs) model produces an exceptionally hard CR proton spectrum peaking at or above the CR knee (i.e., Fig.~\ref{fig:spec}). 
A key constraint on the fraction of ultra-high-energy CRs produced by CSFs in galactic star clusters is the anisotropy. 
To calculate the anisotropy, we have used a simple model for the galactic geometry, interstellar magnetic turbulence, and distribution of CSF sources.
The anisotropy depends strongly on the diffusion coefficient, $D(E)$, 
and we have determined $D(E)$ with a direct calculation of CR trajectories in simulated turbulence (Fig.~\ref{fig:diff1}).
The model described above considered high energy CR  production by a subset of galactic SNe exploding in compact clusters of young massive stars, i.e., CSF systems. 
The mean time between SNe in these CSF systems is taken to be  2000\,yr. The clusters are distributed within the galactic disk as illustrated in Fig.~\ref{fig:clusters}. Cosmic rays are propagated until they leave the thick galactic disk.  

We estimated the ratio of the high energy CR flux produced  by young massive clusters to the total flux needed to provide the observed anisotropy and found that $30-50\%$ of the total at 
$\sim 100$\,PeV may come from clusters without violating observed anisotropy levels.

\section{Acknowledgements}
A.B. and M.K. were supported in part by the RSF grant 16-12-10225. Some of the modeling was
performed at the ``Tornado'' subsystem of the St.~Petersburg
Polytechnic University supercomputing center and at the JSCC RAS.

\section*{References}
\bibliographystyle{aps-nameyear3}
\bibliographystyle{harvard}

\end{document}